\documentclass[aps,prl,preprint]{revtex4-1}
\usepackage{amsfonts}
\usepackage{amsmath}
\usepackage{amssymb}
\usepackage{graphicx}
\usepackage{setspace}
\usepackage{CJK}

\begin{document}

\title{On the analytical formulation of classical electromagnetic fields}
\author{Mark Robert Baker}
\affiliation{Department of Physics and Astronomy, The University of Western Ontario, London, ON, Canada N6A 3K7}
\email{mbaker66@uwo.ca}
\date{\today}
\begin{abstract}
Three objections to the canonical analytical treatment of covariant electromagnetic theory are presented: (i) only half of Maxwell's equations are present upon variation of the fundamental Lagrangian; (ii) the trace of the canonical energy-momentum tensor is not equivalent to the trace of the observed energy-momentum tensor; (iii) the Belinfante symmetrization procedure exists separate from the analytical approach in order to obtain the known observed result. It is shown that the analytical construction from Noether's theorem is based on manipulations that were developed to obtain the compact forms of the theory presented by Minkowski and Einstein; presentations which were developed before the existence of Noether's theorem. By reformulating the fundamental Lagrangian, all of the objections are simultaneously relieved. Variation of the proposed Lagrangian yields the complete set of Maxwell's equations in the Euler-Lagrange equation of motion, and the observed energy-momentum tensor directly follows from Noether's theorem. Previously unavailable symmetries and identities that follow naturally from this procedure are also discussed.
\end{abstract}
 

\maketitle

\section{1. Historical Context}

Electromagnetic theory was developed in large part due to the experimental conclusions of Michael Faraday in the early 1830's, followed by the mathematical reasoning of James Clerk Maxwell in the 1850's. Maxwell first presented his equations in components in 1861 \cite{maxwell1861}, which are unlike what is presented in the modern day. The common form of Maxwell's equations were introduced by Heaviside in 1894, which concisely represents the laws as divergences and curls of electric and magnetic field vectors \cite{heaviside1894}. Parallel developments during the 19th century were occurring in the area of analytical mechanics introduced by Joseph-Louis Lagrange. He showed how experimentally determined equations of motion could be derived from the principle of least action.\\

An obvious curiosity followed; for an experimentally constructed theory like electromagnetism, does an appropriate Lagrangian exist such that the calculus of variations will lead to Maxwell's equations as the equation of motion? The path to such goal was initiated by Hermann Minkowski in 1908 \cite{minkowski1909sr}. In this paper, he noticed that Maxwell's equation could be represented as the divergence of two antisymmetric matrices whose components are that of the electric and magnetic field. He also noticed that one these matrices can be multiplied together, and combined the trace of the fundamental Lagrangian $\mathcal{L} = - \frac{1}{4} {F}_{\lambda\gamma} F^{\lambda\gamma}$, to recover the energy and momentum expressions of the theory. These expressions are the well known energy-density $\frac{1}{2}(E^2 + B^2)$, Poynting's vector $\vec{E} \times \vec{B}$, and the Maxwell stress tensor \cite{maxwell1873}.\\

Einstein, whose love for electrodynamics is well known, is responsible for the current form of the covariant Maxwell's equations, which is not so well known \cite{einstein1916em}. He decided the need for the two fundamental matrices, now shown to transform as true tensors, is not the simplest formulation of the covariant theory. Instead he proposed to drop the second tensor from the formulation, since all of Maxwell's equations could be expressed as the divergence, and Bianchi identity, of the field strength tensor $F_{\mu\nu}$ that is associated with the non-homogenous Maxwell's equations. Furthermore, the energy momentum tensor $T^{\mu\nu}$ could be expressed using only this tensor and the trace of the Lagrangian composed of this tensor, as shown by Minkowski.\\

This form of covariant electrodynamics that has held for a century now. Countless publications have made use of this canonical formulation, and continue to do so in the present day. Most standard electrodynamics textbooks such as Jackson \cite{jackson1998} conclude with this presentation, and in recent years it continues to be the subject of investigation, for example in 2009 \cite{heras2009} and in 2016 \cite{carrasco2016}. It is absolutely satisfying in terms of results, as all of Maxwell's equations, and the energy-momentum expressions, can be expressed compactly using the field strength tensor. The compact form does not, however, guarantee it as a derivable theory of analytical mechanics.\\

Several inconsistencies can be found between the analytical formulation and traditional formulation of electrodynamics. Traditionally, Maxwell's equations can be represented as a pair of wave equations in terms of the electric and magnetic fields \cite{tepper2011}. In the analytical approach, only a single wave equation (Gauss-Ampere) is present in the Euler-Lagrange equation of motion. Analytically, the full symmetry of Maxwell's equations are prohibited, and the source wave equation is given priority. Why should the opinion of physicists determine which equation is obtained upon variation? This question is the basis for objection (i), as explained in section 4.1. The remaining inconsistencies are related to Noether's theorem.\\

In 1918, Emmy Noether proposed an extension of the Lagrangian equation of motion in analytical mechanics, such that the previously discarded boundary term was in fact a conservation law \cite{noether1918}. This conserved quantity is based on symmetry which is associated to the theory, such as Lorentz symmetry. It is remarkable that variation of a fundamental Lagrangian could lead directly to the equation of motion and conservation laws of a theory by following strict rules of procedure. Conventional wisdom states that Noether's theorem derives electromagnetic theory from the fundamental Lagrangian $\mathcal{L} = - \frac{1}{4} {F}_{\lambda\gamma} F^{\lambda\gamma}$ in its entirety. This is not the case, as explained in section 4.2 (objection (ii)) and section 4.3 (objection (iii)). \\

The energy-momentum expressions in electromagnetism are the subject of ongoing investigation \cite{kholmetskii2016}. Electromagnetism from an analytical perspective is less established than recent presentation would indicate \cite{charap2011}. The canonical energy-momentum tensor does not directly give known results in electromagnetic theory. Objection (ii) explains how the canonical energy-momentum tensor has a trace which is not equivalent to the observed energy-momentum tensor. This objection arises in part due to two different presentations of Noether's theorem in the literature, described in section 3.1 and 3.2.  Objection (iii) describes how the symmetrization method exists independent of the analytical approach in order to obtain the known energy-momentum tensor \cite{schust2005}. The current work provides a solution to the three objections which allows for electromagnetism to be obtained directly from Noether's theorem. \\

\section{2. The canonical equations}

Canonical equations of covariant electrodynamics will be presented with $c = 1$ and metric signature $(+, -, -, -)$. Maxwell's equations in free space can be expressed as the Gauss-Ampere law $\vec{\nabla} \cdot \vec{E} = 0$, $\vec{\nabla} \times \vec{B} = \partial_t \vec{E}$, and Gauss-Faraday law $\vec{\nabla} \cdot \vec{B} = 0$, $\vec{\nabla} \times \vec{E} = - \partial_t \vec{B}$. These equations can be solved for the components of the electric and magnetic fields, $\vec{E} = (E_x, E_y, E_z)$ and $\vec{B} = (B_x, B_y, B_z)$, respectively. Einstein compactly defined Maxwell's equations \cite{einstein1916em} in terms of the divergence (Gauss-Ampere) and Bianchi identity (Gauss-Faraday) of the field strength tensor $F_{\mu\nu} = \partial_\mu A_\nu - \partial_\nu A_\mu$,

\begin{equation}
\begin{aligned}
  \partial_\mu F^{\mu\nu} &= 0, \\
  \partial_\alpha F_{\mu\nu} + \partial_\mu F_{\nu\alpha} + \partial_\nu F_{\alpha\mu} &= 0, \label{MaxEqns}
\end{aligned}
\end{equation}

where ${A}^\nu = (\Phi, \vec{A})$ is the four-potential, of the electric scalar potential $\Phi$, and magnetic vector potential $\vec{A}$. The covariant 4-divergence is given as $\partial_\mu = (\partial_t, \partial_x, \partial_y, \partial_z)$. The 6 independent components of this antisymmetric tensor are that of the electric and magnetic field \cite{jackson1998}. The symmetric energy-momentum tensor is given by,

\begin{equation}
T^{\mu\nu} = {F}^{\mu\alpha} F^\nu_{\ \ \alpha} - \frac{1}{4} \eta^{\mu\nu} {F}_{\lambda\gamma} F^{\lambda\gamma} \label{ETuv}
 \end{equation}

The components of this tensor are the energy density of the electromagnetic field, Poynting's vector, and the Maxwell stress tensor, which were all well-established quantities before the covariant formulation was introduced. Differentiation can show on shell energy-momentum conservation, and the force density of the Lorentz force. The covariant formulation above, as developed by Minkowski \cite{minkowski1909sr} and Einstein \cite{einstein1916em}, holds to this day.

\section{3. Analytical treatment of classical electromagnetism} \label{AnalyticalTreatment}

It is of extreme importance to differentiate between two presentations of Noether's theorem found in the literature. First is the presentation following from the principle of least action $\delta S = 0$, for example in \cite{jackiw1994}. This method follows the procedure of Lagrange, to minimize the variation of the action. Instead of neglecting the boundary term from this method, it is treated as the divergence which correspondes to the conserved quantity. This method correspondes to the first section of Noether's paper. Only by manipulations of the $\delta \mathcal{L}$ term such as in \cite{jackiw1994} is this trace subtracted. This subtraction is a desired property in the canonical presentation of analytical electrodynamics, because it is required to obtain the known result. It will be shown that this method can be of interest in constructing electrodynamics analytically. This method is shown in section 3.1.\\

The second method often presented is that of invariant action leading to conservation laws, taken from the second section of Noether's paper. This method is more true to what Noether was emphasizing in \cite{noether1918}. The principle of least action is not what is of interest; instead it is required that two actions are equivalent (form invariant) under a coordinate transformation. In this presentation (see \cite{sundermeyer1982}) the trace term arises more naturally in the canonical energy-momentum tensor from consolidating the two actions under one coordinate variable. This method is not as direct as the least action principle; it requires truncated expansions of the coordinate variation and new definitions of potential variations. However, it does provide a proper procedure for obtaining the trace substraction required in the canonical analytical formulation of electrodynamics. This method is shown in section 3.2.

\subsection{3.1. Electrodynamics from least action principle}

From the principle of least action $\delta S = 0$, the variation of the Lagrangian function must be minimized in order to obtain an equation of motion. The minimal variation of a Lagrangian built from a vector potential and its derivatives is,

\begin{equation}
\delta \mathcal{L} = \frac{\partial \mathcal{L}}{\partial (A_\alpha)} \delta A_\alpha + \frac{\partial \mathcal{L}}{\partial (\partial_\mu A_\alpha)} \partial_\mu \delta A_\alpha + \frac{\partial \mathcal{L}}{\partial (\partial_\mu \partial_\nu A_\alpha)} \partial_\mu \partial_\nu \delta A_\alpha + ... = 0.
\end{equation}

No second order or higher derivatives of potential are present in the Lagrangian of classical electrodynamics so they can be neglected. Performing integration by parts yields,

\begin{equation}
\delta \mathcal{L} = [\frac{\partial \mathcal{L}}{\partial (A_\alpha)}  -  \partial_\mu \frac{\partial \mathcal{L}}{\partial (\partial_\mu A_\alpha)}]  \delta A_\alpha + \partial_\mu [\frac{\partial \mathcal{L}}{\partial (\partial_\mu A_\alpha)}  \delta A_\alpha] = 0, \label{Noether}
\end{equation}

where the first term forms the Euler-Lagrange equation, and the second term is the boundary term which Lagrange discarded from the action. In order to analytically treat electromagnetic theory, the fundamental electromagnetic Lagrangian $\mathcal{L} = - \frac{1}{4} {F}_{\lambda\gamma} F^{\lambda\gamma}$ must be varied in order to obtain an equation of motion. The boundary term, which yields the canonical energy-momentum tensor, will not be discarded in Noether's theorem. Variation yields the equation of motion and conserved quantity,

\begin{equation}
\begin{aligned}
  \partial_\mu F^{\mu\nu} &= 0, \\
  \partial_\mu [F^{\mu\nu} \delta A_\nu] &= 0. \label{ConQuan}
\end{aligned}
\end{equation}

The equation of motion, $ \partial_\mu F^{\mu\nu}$, is only half of Maxwell's equations (Gauss-Ampere law). This is immediately obvious since this equation is a 4-vector, and 8 equations exist in the complete set of Maxwell's equations (equation \ref{MaxEqns}). The Gauss-Faraday law exists separate of the analytical approach. If only half of the equation of motion is following from variation, can this formulation be considered theoretically complete? This is the basis for objection (i), described in section 4.1.\\

Taking the variation of the potential under a Lorentz translation ($x_\nu \rightarrow x_\nu + \delta x_\nu$) to be $\delta A_\alpha =   \partial^\nu A_\alpha \delta x_\nu$, and the variation of the Lagrangian under a Lorentz translation to be $\delta \mathcal{L} = \partial_\mu ( \eta^{\mu \nu} \mathcal{L} \delta x_{\nu})$, the conserved quantity can be expressed as a second rank tensor by removing the coordinate variation (which is taken to be a constant) from the expression. It is important to note that the $\delta \mathcal{L}$ term is taken from the left hand side of the Lagrangian variation in order to combine with the trace and obtain the energy-momentum tensor $T^{\mu\nu}$ of Minkowski and Einstein,

\begin{equation}
\delta \mathcal{L} =  \partial_\mu [\frac{\partial \mathcal{L}}{\partial (\partial_\mu A_\alpha)}  \delta A_\alpha] = 0. \label{Objiieqn}
 \end{equation}

This is the basis for the objection (ii), described in section 4.2. It is clear from equation \ref{ConQuan} that the second term in equation \ref{ETuv} is not directly following from least action. This term is created from the Lagrangian variation (equation \ref{Objiieqn}). It  is imposed similar to the Belinfante procedure, in order to obtain known results in electrodynamics that existed before Noether's theorem. This is how the conserved quantity becomes the canonical energy-momentum tensor,

\begin{equation}
\partial_\mu T^{\mu\nu} = \partial_\mu [F^{\mu\alpha} \partial^\nu A_\alpha + \eta^{\mu \nu} \mathcal{L}] = 0. \label{TuvC}
 \end{equation}

Based on the Lorentz translation symmetry, the electromagnetic Lagrangian gives rise to the so called canonical energy-momentum tensor $T_C^{\mu\nu} = F^{\mu\alpha} \partial^\nu A_\alpha - \frac{1}{4} \eta^{\mu \nu} {F}_{\lambda\gamma} F^{\lambda\gamma}$. This is not in the form of the observed, symmetric expression presented by Minkowski and Einstein. In 1940, Frederik Belinfante proposed a solution, by defining a symmetrization procedure \cite{belinfante1940}. The canonical energy-momentum tensor combined with the Belinfante correction term $\partial_\alpha b^{\mu \nu \alpha} = - F^{\mu\alpha} \partial_\alpha A^\nu$ yielded the expression of Einstein (equation \ref{ETuv}). Basically, Belinfante used the fact that any tensor can be expressed as the combination of symmetric and antisymmetric parts to determine what was missing from the symmetric form. After the correction was made, this formulation of electromagnetic theory has remained untouched until present day. \\

\subsection{3.2. Electrodynamics from invariant actions}

The second common presentation is that of invariant actions, where two actions are said to be form invariant under a coordinate transformation. In this case the action as a function of the potential is equivalent to the same action as a function of the transformed potential, $S(A_\mu) = S(\hat{A}_\mu)$. Requiring form invariance leads to the following condition in the action,

\begin{equation}
\frac{\partial \hat{x_\alpha}}{\partial x_\alpha} \mathcal{L}(\hat{A}_\mu) - \mathcal{L}(A_\mu) = 0 .
 \end{equation}

Taking the truncated expansion of the coordinate variation (first order approximation), the first term can be expressed as $\frac{\partial \hat{x_\alpha}}{\partial x_\alpha} \approx 1 + \partial^\alpha (\delta x_\alpha)$. Such requirements take away from the absolute purity of the least action procedure. Here the variation of the Lagrangian is not imposed, rather it is defined as a consequence of the equality $\delta \mathcal{L} = \mathcal{L} (\hat{A}_\mu) - \mathcal{L} (A_\mu)$. Therefore the form invariance requirement can be expressed as,

\begin{equation}
\delta \mathcal{L} + \mathcal{L}(\hat{A}_\mu) [\partial^\alpha (\delta x_\alpha)]  = 0 ,
 \end{equation}
 
which can also be expressed as $\delta \mathcal{L} + \partial^\alpha [\mathcal{L}(\hat{A}_\mu) \delta x_\alpha] - [\partial^\alpha \mathcal{L}(\hat{A}_\mu)]  \delta x_\alpha = 0$. The second (divergence) term is the origin of the trace subtraction in the canonical energy-momentum tensor. Consolidating the first and third terms under a change of variables (noted as bar $\bar{\delta} A_\beta = \hat{A}_\mu - A_\mu$) yields the conserved quantities,

\begin{equation}
\partial^\alpha [\frac{\partial \mathcal{L}}{\partial (\partial_\alpha A_\beta)} \bar{\delta} A_\beta] + \partial^\alpha [\mathcal{L}(\hat{A}_\mu) \delta x_\alpha] = 0. \label{invariantnoether}
 \end{equation}
 
 Here the divergence expression from equation \ref{Objiieqn} follows from the variation of the Lagrangian, and a term which can be expressed as the trace subtraction is also present. Both the least action principle, and invariant action requirement have been tailored to obtain the same canonical energy-momentum tensor used to construct electrodynamics. It will be shown in section 5 that regardless of the approach, the proposed reformulation recovers completely the equations of electrodynamics. In both approaches, the Belinfante procedure is required to obtain the energy-momentum tensor.\\

The fact that the Belinfante procedure exists separately from variational methods is the basis for objection (iii), described in section 4.3. The fundamental Lagrangian is not directly yielding the observed energy-momentum tensor in Noether's theorem. In order to make this be the case, the Belinfante procedure adds a term to what is following from Noether's theorem. This is an ad-hoc addition in order to obtain known results. Furthermore, the name 'symmetrization procedure' is misleading. There is no evidence that symmetry is a fundamental property of the energy-momentum tensor, symmetrization just coincidentally yields the form presented by Minkowski and Einstein.

\section{4. Objections to the canonical treatment}

\subsection{4.1. Only half of Maxwell's equations follow directly from variation of the fundamental Lagrangian $\mathcal{L} = - \frac{1}{4} {F}_{\lambda\gamma} F^{\lambda\gamma}$} \label{Obj1}

This objection, perhaps most obvious, is one that is held by weak arguments in typical textbook and literature presentation \cite{jackson1998} \cite{misner1973gravitation}. Since only half of Maxwell's equations follow from Noether's theorem, the other half must exist separately from the analytical approach. Analytical methods are not held in such high regards in all circles, and most are satisfied simply with the compact form of covariant electromagnetism, as was Einstein. Even though it is obvious only half of the equations are present in the Euler-Lagrange equation, perhaps due to tradition and the name of Einstein, the Bianchi identity existing independent of variational methods has been considered to be sufficient. It is important to realize that the current covariant equations were developed before Noether's theorem. Analytically obtaining electromagnetic theory from Noether's theorem has been based strictly on obtaining the covariant equations of Minkowski and Einstein.\\

While the fact that the Euler-Lagrange equation does not yield the complete set of equations from the fundamental Lagrangian $\mathcal{L} = - \frac{1}{4} {F}_{\lambda\gamma} F^{\lambda\gamma}$ should be suspicious enough to some, a more compelling argument is the concept of bias. The opinion of a theoretical physicist should not enter the determination of a theory. Here, the Gauss-Ampere law is given preferential treatment due to the existence of electric charge. Other than this, there is no logical reasoning for the preferential treatment. Maxwell's Gauss-Ampere and Gauss-Faraday laws are absolutely equivalent in free space. The canonical field strength $F_{\mu\nu}$ is simply which was chosen by Einstein to represent the theory in compact form. \\

Just as easily, the field strength representing the Gauss-Faraday law, as defined by Minkowski, could be used to build the entire theory. Minkowski called this the dual field strength $\mathcal{F}_{\mu\nu}$, and it is well known to give the second half of Maxwell's equations, independent of analytical methods. If electromagnetism is built analytically from the dual perspective Lagrangian $\mathcal{L} = - \frac{1}{4} \mathcal{F}_{\lambda\gamma} \mathcal{F}^{\lambda\gamma}$, the Gauss-Faraday law $\partial_\mu \mathcal{F}^{\mu\nu} = 0$ will exist in the equation of motion, and Gauss-Ampere law as the Bianchi identity $\partial_\alpha \mathcal{F}_{\mu\nu} + \partial_\mu \mathcal{F}_{\nu\alpha} + \partial_\nu \mathcal{F}_{\alpha\mu} = 0$. Energy-momentum tensor $T^{\mu \nu} = \mathcal{F}^{\mu\alpha} \mathcal{F}^\nu_{\ \ \alpha} - \frac{1}{4} \eta^{\mu\nu} \mathcal{F}_{\lambda\gamma} \mathcal{F}^{\lambda\gamma}$ is exactly the observed tensor given in equation \ref{ETuv}. The entire theory is present in compact form. Why should the convention of theoretical physicists determine a fundamental Lagrangian? More importantly, why should we be satisfied by a Lagrangian which does not yield a complete theory upon variation? This bias will be central to the development of this work, as well as the concept of duality and invariant Lagrangian construction.

\subsection{4.2. The trace of the conserved quantity is not what is observed} \label{Obj2}

The second objection is related to the feature noticed by Minkowski and Einstein that contraction of the field strength tensor to form a symmetric second rank tensor ${F}^{\mu\alpha} F^\nu_{\ \ \alpha}$ differs from the observed energy-momentum tensor (equation \ref{ETuv}) by only the trace. What they noticed is that by combining the fundamental Lagrangian $\mathcal{L} = - \frac{1}{4} {F}_{\lambda\gamma} F^{\lambda\gamma}$ with the trace of this combination, the observed energy-momentum tensor developed by Minkowski could be obtained. This is why the canonical energy-momentum expression in equation \ref{TuvC} includes the second term. It is only included to obtain the known result of Minkowski, who died before Noether's theorem was developed. Once again, this energy-momentum tensor was introduced as a compact, covariant notation. It was introduced before Noether's theorem was developed.  \\

Section 3.1. and 3.2. describe two approaches commonly presented as Noether's theorem which are in fact quite different. The first uses the traditional least action principle $\delta S = 0$. As seen in equation \ref{TuvC}, the energy-momentum expression does not follow directly from Noether's theorem. Since no subtraction of trace exists in the conserved quantity (equation \ref{ConQuan}), the variation expression $\delta \mathcal{L}$ was manipulated in order to obtain this term. An explicit account of this process is given in \cite{jackiw1994}. In fact, is not true to the least action principle, since this term is identically zero (equation \ref{Noether}). It is argued that $\delta \mathcal{L} = \partial_\mu ( \eta^{\mu \nu} \mathcal{L} \delta x_{\nu})$, in order to pull this term across and into the conserved quantity. There is no physical or mathematical reasoning to do this, other that it can be used to obtain the known result. Acceptance of this process is based entirely on correctly obtaining the energy-momentum tensor in the form of Minkowski and Einstein.\\

The second approach requires two actions under a coordinate transformation to be form invariant. In this approach two divergence terms can be obtained: one which is the divergence as in the least action principle, and another which can be expressed as the trace subtraction. This method requires change of variables and approximations which take away from the exact nature of the least action principle, however can give a more mathematical procedure if one desires the trace subtraction term. It is shown in section 5 that regardless of which method is chosen, the reformulation will lead uniquely to the equations of electrodynamics.

\subsection{4.3. The Belinfante symmetrization procedure of the canonical energy-momentum tensor} \label{Obj3}

The third and final objection is another manipulation of the canonical energy-momentum tensor $T^{\mu\nu}_C$. Although this procedure yields the desired result for electromagnetism, it exists completely separate of variational methods, and is only accepted because it manipulates the variational result into the known $T^{\mu\nu}$ expression of Minkowski and Einstein. Since the observed energy-momentum expression does not follow from Noether, this addition is necessary to obtain the desired result. The procedure cares only about the symmetrization of an arbitrary tensor by adding a correction term $\partial_\alpha b^{\mu \nu \alpha} = - F^{\mu\alpha} \partial_\alpha A^\nu$, which has no physical connection. The fact that symmetrizing Noether's conserved quantity leads to the observed energy-momentum tensor is a coincidence. The Belinfante procedure is simply a way for making an arbitrary tensor symmetric. Such manipulation can in no way be considered fundamental to analytical mechanics. Unfortunately, this is often presented as some natural step in the derivation of electromagnetic theory from analytical approaches, and few instances can be found where serious discussion has taken place. In 1980, a more theoretically sound potential variation was introduced by considering the potential transformation under both Lorentz and gauge symmetries by Eriksen and Leinhaas \cite{eriksen1980}. A clear presentation is given in the classical field theory book of Burgess \cite{burgess2002}, which will be discussed in section 5. \\

One can raise the question after reading these objections, if there are problems with the equation of motion, and the conservation law, is anything following naturally from Noether's theorem in electromagnetic theory? Furthermore, if the complete set of Maxwell's equations are not present in the equation of motion, how can the incomplete theory be expected to yield the full energy-momentum tensor in the first place? After all, the energy-momentum tensor was originally found from the complete set of equations. Perhaps the problem behind these objections is one in the same.

\section{5. The solution to the objections}

The purpose of this work is not only to address the objections to the canonical treatment of covariant electrodynamics, but to offer a solution that is superior to current formulation. The solution lies at the construction of the Lagrangian itself. As described in section 3, the choice of field strength is rather arbitrary, and the theory can be equivalently expressed in term of the dual field strength of Minkowski \cite{minkowski1909sr}. This dual field strength $\mathcal{F}_{\mu\nu}$ is again defined in terms of the field strength $F_{\mu\nu}$ via contraction of the Levi-Civita tensor $\mathcal{F}_{\mu\nu} = \epsilon_{\mu\nu\alpha\beta} F^{\alpha \beta}$ \cite{jackson1998} \cite{heras2009}. This relation is not advantageous, as it continues to prioritize half of Maxwell's equations defined by $F_{\mu\nu}$, and cannot be used in a fundamental Lagrangian because it is not invariant under a parity transformation.\\

Another option remains, which is to define the Gauss-Faraday law with dual field strength $\mathcal{F}_{\mu\nu} = \partial_\mu \mathcal{A}_\nu - \partial_\nu \mathcal{A}_\mu$ under its own (dual) vector potential $\mathcal{A}^\nu = (\Phi_B, \vec{A}_E)$. Here the time component corresponds to an magnetic scalar potential $\Phi_B$, and the space components to an electric vector potential $\vec{A}_E$, which are introduced to the Gauss-Ampere equations. Introducing a second fundamental vector potential may seem like a step in the wrong direction, but this was the original formulation of Minkowski, only removed by Einstein due to a perceived simplification of the compact form. Minkowski treated these like independent quantites. It is clear that this is a natural way of obtaining all 8 of Maxwell's equations, since the single 4-vector will only yield 4 equations upon varition. Allowing for variation of a second 4-dimensional vector potential results in an additional 4 equations in the Euler-Lagrange expression, which is identically half of Maxwell's equations.\\

Under the invariant $\mathcal{F}_{\lambda\gamma} \mathcal{F}^{\lambda\gamma}$, electromagnetic theory can be entirely represented in the opposite situation to the traditional approach, where the Euler-Lagrange equation is the Gauss-Faraday law $\partial_{\lambda} \mathcal{F}^{\lambda\gamma}$, and the Bianchi identity recovers the Gauss-Ampere law. The energy-momentum tensor $T^{\mu \nu} = \mathcal{F}^{\mu\alpha} \mathcal{F}^\nu_{\ \ \alpha} - \frac{1}{4} \eta^{\mu\nu} \mathcal{F}_{\lambda\gamma} \mathcal{F}^{\lambda\gamma}$ can also be recovered from manipulation and the Belinfante procedure. Fundamental Lagrangian $\mathcal{L} = - \frac{1}{4} {F}_{\lambda\gamma} {F}^{\lambda\gamma}$ is clearly not special in this regard. It is curious to wonder how to select which is truly more fundamental. Electromagnetic theory does not follow without manipulation in either individual situation. After lengthy consideration, it became clear that both invariants must be included to avoid bias, and recover the entire theory without manipulations. For the following calculations it is useful to work with the field strength tensors explicitly in terms of the components of the electric and magnetic fields, \\

\begin{equation}
F_{\mu\nu} = 
 \begin{pmatrix}
  0 & E_x & E_y & E_z \\
  - E_x & 0 & -B_z & B_y \\
  - E_y  & B_z  & 0 & -B_x  \\
  - E_z & - B_y & B_x & 0 
 \end{pmatrix},
\quad
\mathcal{F}_{\mu\nu} = 
 \begin{pmatrix}
  0 & B_x & B_y & B_z \\
  - B_x & 0 & E_z & -E_y \\
  - B_y  & -E_z  & 0 & E_x  \\
  - B_z & E_y & -E_x & 0 
 \end{pmatrix}.
 \end{equation}

It is noted that other possible scalars ${F}_{\lambda\gamma} \mathcal{F}^{\lambda\gamma} = \mathcal{F}_{\lambda\gamma} {F}^{\lambda\gamma}$ were seriously considered. However, they simulatenously destroy the symmetry of the equation of motion, energy-momentum tensor, conservation identities, and do not appear to be fundamentally significant from an analytical approach. This is related to the fact that under the Levi-Civita dual treatment, this combination is only a pseudoscalar, thus cannot be part of a Lorentz invariant field theory. \\

Natural progression of logic leads to the desire for a fundamental Lagrangian such that the theory is directly following from Noether's theorem. With several available objections to the previous treatment, attributing these problems individually to the variational method itself is extremely unlikely. By defining a Lagrangian that is a combination of these,

\begin{equation} 
\mathcal{L} = - \frac{1}{4}[{F}_{\lambda\gamma} {F}^{\lambda\gamma} + \mathcal{F}_{\lambda\gamma} \mathcal{F}^{\lambda\gamma}],
\end{equation}

 all objections can simultaneously be erased. The beauty of this approach is hard to deny, and it is proposed as the fundamental Lagrangian to electromagnetic theory. Performing variation with respect to both potentials yields,

  \begin{equation}
 \delta \mathcal{L} =  [\partial_\mu  {F}^{\mu\nu} ] \delta A_\nu + [\partial_\mu \mathcal{F}^{\mu \nu}] \delta \mathcal{A}_\nu  + \partial_\mu [-{F}^{\mu\alpha} \delta A_\alpha - \mathcal{F}^{\mu \alpha} \delta \mathcal{A}_\alpha].
\end{equation}

It is clear that this method produces the entire set of Maxwell's equations in the Euler-Lagrange equations of motion, without the need for bias or preferential treatment. Objection 4.1 is relieved, with Gauss-Ampere law $\partial_\mu  {F}^{\mu\nu} = 0$ and Gauss-Faraday law $\partial_\mu \mathcal{F}^{\mu\nu} = 0$. The Biachi identity can even be implemented here if the equations are again desired in terms of one of the field strength tensors. \\

This Lagrangian also has a unique property that $\mathcal{L} = B^2 - E^2 - B^2 + E^2 = 0$. Therefore in the trace subtraction term of the canonical energy-momentum tensor where $\mathcal{L}$ exists separately (equation \ref{TuvC} and equation \ref{invariantnoether}), this term independently vanishes. Due to this property, using the method of least action principle or invariant action will lead to the same result. This indicates that both Noether presentations of section 3.1 and 3.2 are compatable for very specific Lagrangians.\\

The proper variation of the electromagnetic potential vectors was well defined by Eriksen and Leinhaas \cite{eriksen1980} \cite{burgess2002}. It was shown that gauge invariant potential variation naturally follows under the simultaneous consideration of both Lorentz and gauge symmetries of electrodynamic theory. This work has gone relatively unnoticed, but answered two physical questions about potential variation as required for the symmetry of Noether's theorem, 

\begin{equation}
\begin{aligned}
  \delta A_\alpha = A'_\alpha - A_\alpha = \delta x_\nu F^\nu_{\ \ \alpha}&, \\
  \delta \mathcal{A}_\alpha = \mathcal{A}'_\alpha - \mathcal{A}_\alpha = \delta x_\nu \mathcal{F}^\nu_{\ \ \alpha}&.
\end{aligned}
\end{equation}

 First, it showed that the potential difference, which is observable, is indeed gauge invariant. Second, it can be used to recover known expressions for potential differences (i.e. voltages), such as $\delta x_2 F^2_{\ \ 1} = d_x E_x = V' - V = \Delta V$. It is well known that gauge symmetries inherent to electrodynamics. Why should potential variation therefore only depend on Lorentz symmetries? The dramatic improvement of the analytical approach from this variation warrants a serious rethinking of classical potential variation. This result is the true solution to the Belinfante problem, so the original authors Eriksen and Leinhaas deserve credit where it is due. \\

The true beauty of the reformulation, however, is the fact that the energy-momentum tensor is,

        \begin{equation}
T^{\mu\nu} =  \frac{1}{2}( {F}^{\mu\alpha} F^\nu_{\ \ \alpha} +  \mathcal{F}^{\mu\alpha} \mathcal{F}^\nu_{\ \ \alpha}) =   \begin{pmatrix}
- \frac{1}{2}(E^2 + B^2) & - S_x & - S_y & - S_z \\
  -S_x  & \sigma_{xx} & \sigma_{xy} & \sigma_{xz} \\
  -S_y   & \sigma_{yx}  & \sigma_{yy} & \sigma_{yz}  \\
  -S_z  & \sigma_{zx} &\sigma_{zy} & \sigma_{zz}
 \end{pmatrix},
 \end{equation}

which is exactly equivalent to the well known energy-momentum tensor presented by Minkowski and Einstein in equation \ref{ETuv}, where $\vec{S}$ is Poynting's vector, and components $\sigma_{ij}$ form the Maxwell stress-tensor. This energy-momentum tensor is identical to that of Minkowski in equation \ref{ETuv}, without the need for ad-hoc additions to the analytical approach. Instead of creating a term to subtract from the trace, the combination of the dual field strength serves a similar purpose. This is because these two contractions (field strength and dual) only differ on the trace. While the tensors that build this energy-momentum tensor are different than the traditional approach, all energy-momentum laws in electrodynamic theory are again present in the exact same form. Therefore objection 4.2 and objection 4.3 have also been relieved.  \\

 The equations of motion can then be used to show on shell conservation $\partial_\mu T^{\mu\nu} = 0$. What is interesting is that in this approach the Bianchi identity is not required, as this information is present in the dual formulation. Instead, a useful identity ${F}^{\mu\alpha} [\partial_\mu F_{\omega\alpha}]  + \mathcal{F}^{\mu\alpha} [\partial_\mu \mathcal{F}_{\omega \alpha}] = 0$ can be found by breaking the remaining terms into components. The perfect combination of these components is not obvious, as the equations of motion exist only through the mixing of these two terms. Mixing of terms in this identity highlights the symmetric nature of electrodynamics that naturally follows from the proposed Lagrangian.\\
 
 Furthermore, under Lorentz rotation $\delta A_\alpha = \omega_{\lambda \gamma} X^\gamma F^\lambda_{\ \ \alpha}$, the angular momentum tensor $M^{\mu\nu\lambda} = T^{\mu\nu} X^\lambda  - T^{\mu \lambda} X^\nu$ \cite{jackson1998} directly follows from the conserved quantity  $\omega_{\lambda \nu} \partial_\mu [-{F}^{\mu\alpha} X^\nu F^\lambda_{\ \ \alpha} + {F}^{\mu\alpha} X^\lambda F^\nu_{\ \ \alpha} - \mathcal{F}^{\mu \alpha} X^\nu \mathcal{F}^\lambda_{\ \ \alpha} + \mathcal{F}^{\mu \alpha} X^\lambda \mathcal{F}^\nu_{\ \ \alpha}]$ in Noether's theorem. The fact that the entire theory follows from the reformulation is evidence enough to consider its validity. Above this, the previous objections are no longer an issue, and Maxwell's theory of electromagnetic fields can follow directly from analytical mechanics, without the need for manipulations to obtain known results.
 
 \section{6. Charge, current and Lorentz force}
 
 The introduction of charge and current to the reformulation highlights a major benefit of considering the proposed Lagrangian to be fundamental. In the prior treatment, the coupled $A_\alpha J^\alpha = A_\alpha J_E^\alpha$ electric monopole term yielded the source for the Gauss-Ampere wave equation. This method is not entirely symmetric because it restricts the possibility of a magnetic charge and current density for the source of the Gauss-Faraday equations, which exist separate of variational methods. Such restrictions do not exist to the equations in their divergence and curl forms. The coupling of charge and current densities can introduce a source to the appropriate equations, ${A}_\alpha J_E^\alpha$ and $\mathcal{A}_\alpha J_B^\alpha$, yielding equations of motion,

\begin{equation}
\begin{aligned}
  \partial_\mu  {F}^{\mu\nu} &= J_E^\nu, \\
  \partial_\mu \mathcal{F}^{\mu \nu} &= J_B^\nu,
\end{aligned}
\end{equation}

where $J_B^\mu$ is introduced in presence of magnetic monopoles. The aforementioned conservation identity becomes, under the presence of charge and current, 

\begin{equation} 
{F}^{\mu\alpha} [\partial_\mu F_{\omega\alpha}]  + \mathcal{F}^{\mu\alpha} [\partial_\mu \mathcal{F}_{\omega \alpha}] =  {F}_{\omega\alpha} J_E^\alpha + \mathcal{F}_{\omega\alpha} J_B^\alpha.
\end{equation}

These terms are not independently equivalent, they follow from the mixing of the terms when expressed in components. In the canonical theory with no magnetic monopoles, this identically recovers the force density $f^\nu = - \partial_\mu T^{\mu \nu} = \eta^{\nu\beta} F_{\beta \alpha} J_E^\alpha$, which can be used to recover the Lorentz force. Analytically treating the proposed Lagrangian is truly symmetric and all encompassing of electromagnetic theory.\\
 
A note should be made regarding the Lorentz force. This law is expressed in terms of the field strength tensor in the presence of electric monopoles. By introducing the independent dual to the Lagrangian approach, this tensor is also fundamental and allows for the definition of a second Lorentz force, which is required in the presence of magnetic monopoles. Both force densities are recovered by differentiation of the field strength tensor. The Lorentz force is a vital component of the analytical approach, as it should be. By allowing the reformulation, all aspects of electromagnetic field theory are explicitly following from variation of the Lagrangian. Reformulation allows for a much more symmetric presentation of electromagnetic theory from analytical methods, without the need for manipulation to obtain known results.

\section{7. Conclusions}

Three major holes in the canonical presentation of covariant electrodynamics are presented with respect to the analytical treatment. First, only half of Maxwell's equations are present in the analytical treatment of electrodynamics. Second, the canonical energy-momentum tensor has a trace which is not equal to the observed energy-momentum tensor, the second term in equation \ref{ETuv} is created to obtain the known result of Minkowski. Part of this confusion arose from Noether's theorem being presented in two ways: by the princple of least action, and by the requirement of invariant action. Third, the Belinfante symmetrization procedure adds a term to convert the canonical energy-momentum tensor into the observed form. It is shown how these objections prevent the exact theory to be obtained from Noether's theorem. The reason for this long standing problem going unnoticed is because Einstein formulated covariant electrodynamics in a correct compact way, in terms of a single field strength. This was before Noether's theorem. Afterwards, physicists such as Belinfante manipulated the fundamental Lagrangian $\mathcal{L} = - \frac{1}{4} {F}_{\lambda\gamma} F^{\lambda\gamma}$ into obtaining covariant equations of Minkowski and Einstein from Noether's theorem, instead of asking questions about the impurity of the procedure. Since the early 1940s, these methods have been copied from textbook to textbook and in the literature without question. \\

After questioning the holes of the canonical treatment, the appropriate reformulation of the fundamental Lagrangian is proposed as $\mathcal{L} = - \frac{1}{4}[{F}_{\lambda\gamma} {F}^{\lambda\gamma} + \mathcal{F}_{\lambda\gamma} \mathcal{F}^{\lambda\gamma}]$. The dual field strength $\mathcal{F}_{\mu\nu}$ of Minkowski was introduced, not from Levi-Civita contraction $\mathcal{F}_{\mu\nu} = \epsilon_{\mu\nu\alpha\beta} F^{\alpha \beta}$ as often discussed, but from an indepedent dual potential vector $\mathcal{A}_\alpha$. This allows for the inclusion of $\mathcal{F}_{\lambda\gamma} \mathcal{F}^{\lambda\gamma}$ as a fundamental invariant. Einstein moved away from this formulation by defining everything in terms of ${F}_{\mu\nu}$, however it has been shown to be the natural solution to the three objections. \\

The proposed formulation returns covariant electrodynamics to a more similar presentation that was originally introduced by Minkowski. Two field strength tensors are required to represent the full set of Maxwell's equations. Since the symmetry of Maxwell's equations allow for Einstein's formulation in terms of either the field strength tensor ${F}_{\mu\nu}$, or the dual field strength $\mathcal{F}_{\mu\nu}$ tensor, there is no longer the need to select one of these for analytical electrodynamics. Preferential selection of one of these tensors should not be subject to the opinion of physicists. By defining a Lagrangian in terms of both tensors, both invariants are included in the theory, and the observable equations naturally follow without need for bias. One additional strength of this method is that the equations of electrodynamics follow from both the least action principle, and the requirement of invariant action. Resulting from the reformulation is a truly symmetric presentation of electromagnetic theory obtained from analytical mechanics. It has been made clear that compact notation of experimentally verified theories should not be considered as fundamentally sound. Only when these theories can both be experimentally determined, and derived from strict rules of procedure such as those presented by Lagrange and Noether, should they be considered truly significant. \\

\bibliographystyle{apsrev4-1}
\bibliography{CEMBakerBib}

\begin{thebibliography}{19}%
\makeatletter
\providecommand \@ifxundefined [1]{%
 \@ifx{#1\undefined}
}%
\providecommand \@ifnum [1]{%
 \ifnum #1\expandafter \@firstoftwo
 \else \expandafter \@secondoftwo
 \fi
}%
\providecommand \@ifx [1]{%
 \ifx #1\expandafter \@firstoftwo
 \else \expandafter \@secondoftwo
 \fi
}%
\providecommand \natexlab [1]{#1}%
\providecommand \enquote  [1]{``#1''}%
\providecommand \bibnamefont  [1]{#1}%
\providecommand \bibfnamefont [1]{#1}%
\providecommand \citenamefont [1]{#1}%
\providecommand \href@noop [0]{\@secondoftwo}%
\providecommand \href [0]{\begingroup \@sanitize@url \@href}%
\providecommand \@href[1]{\@@startlink{#1}\@@href}%
\providecommand \@@href[1]{\endgroup#1\@@endlink}%
\providecommand \@sanitize@url [0]{\catcode `\\12\catcode `\$12\catcode
  `\&12\catcode `\#12\catcode `\^12\catcode `\_12\catcode `\%12\relax}%
\providecommand \@@startlink[1]{}%
\providecommand \@@endlink[0]{}%
\providecommand \url  [0]{\begingroup\@sanitize@url \@url }%
\providecommand \@url [1]{\endgroup\@href {#1}{\urlprefix }}%
\providecommand \urlprefix  [0]{URL }%
\providecommand \Eprint [0]{\href }%
\providecommand \doibase [0]{http://dx.doi.org/}%
\providecommand \selectlanguage [0]{\@gobble}%
\providecommand \bibinfo  [0]{\@secondoftwo}%
\providecommand \bibfield  [0]{\@secondoftwo}%
\providecommand \translation [1]{[#1]}%
\providecommand \BibitemOpen [0]{}%
\providecommand \bibitemStop [0]{}%
\providecommand \bibitemNoStop [0]{.\EOS\space}%
\providecommand \EOS [0]{\spacefactor3000\relax}%
\providecommand \BibitemShut  [1]{\csname bibitem#1\endcsname}%
\let\auto@bib@innerbib\@empty
\bibitem [{\citenamefont {Maxwell}(1861)}]{maxwell1861}%
  \BibitemOpen
  \bibfield  {author} {\bibinfo {author} {\bibfnamefont {J.~C.}\ \bibnamefont
  {Maxwell}},\ }\href {\doibase 10.1080/14786446108643067} {\bibfield
  {journal} {\bibinfo  {journal} {Philosophical Magazine Series 4}\ }\textbf
  {\bibinfo {volume} {21}},\ \bibinfo {pages} {338} (\bibinfo {year}
  {1861})}\BibitemShut {NoStop}%
\bibitem [{\citenamefont {Heaviside}(1894)}]{heaviside1894}%
  \BibitemOpen
  \bibfield  {author} {\bibinfo {author} {\bibfnamefont {O.}~\bibnamefont
  {Heaviside}},\ }\href@noop {} {\emph {\bibinfo {title} {Electrical
  papers}}},\ Vol.~\bibinfo {volume} {2}\ (\bibinfo {year} {1894})\BibitemShut
  {NoStop}%
\bibitem [{\citenamefont {Minkowski}(1909)}]{minkowski1909sr}%
  \BibitemOpen
  \bibfield  {author} {\bibinfo {author} {\bibfnamefont {H.}~\bibnamefont
  {Minkowski}},\ }\href@noop {} {\bibfield  {journal} {\bibinfo  {journal}
  {Physikalische Zeitschrift}\ }\textbf {\bibinfo {volume} {10}},\ \bibinfo
  {pages} {104} (\bibinfo {year} {1909})}\BibitemShut {NoStop}%
\bibitem [{\citenamefont {Maxwell}(1873)}]{maxwell1873}%
  \BibitemOpen
  \bibfield  {author} {\bibinfo {author} {\bibfnamefont {J.~C.}\ \bibnamefont
  {Maxwell}},\ }\href@noop {} {\emph {\bibinfo {title} {A Treatise on
  Electricity and Magnetism}}},\ Vol.~\bibinfo {volume} {2}\ (\bibinfo {year}
  {1873})\BibitemShut {NoStop}%
\bibitem [{\citenamefont {Einstein}(1916)}]{einstein1916em}%
  \BibitemOpen
  \bibfield  {author} {\bibinfo {author} {\bibfnamefont {A.}~\bibnamefont
  {Einstein}},\ }\href@noop {} {\bibfield  {journal} {\bibinfo  {journal}
  {Sitzungsber. Preuss. Akad. Wiss. Berlin, Math.-Phys Klasse}\ ,\ \bibinfo
  {pages} {184}} (\bibinfo {year} {1916})}\BibitemShut {NoStop}%
\bibitem [{\citenamefont {Jackson}(1998)}]{jackson1998}%
  \BibitemOpen
  \bibfield  {author} {\bibinfo {author} {\bibfnamefont {J.}~\bibnamefont
  {Jackson}},\ }\href {https://books.google.ca/books?id=tYEAkAEACAAJ} {\emph
  {\bibinfo {title} {Classical Electrodynamics}}}\ (\bibinfo  {publisher}
  {Wiley},\ \bibinfo {year} {1998})\BibitemShut {NoStop}%
\bibitem [{\citenamefont {Heras}\ and\ \citenamefont
  {Báez}(2009)}]{heras2009}%
  \BibitemOpen
  \bibfield  {author} {\bibinfo {author} {\bibfnamefont {J.}~\bibnamefont
  {Heras}}\ and\ \bibinfo {author} {\bibfnamefont {G.}~\bibnamefont {Báez}},\
  }\href {http://stacks.iop.org/0143-0807/30/i=1/a=003} {\bibfield  {journal}
  {\bibinfo  {journal} {Eur. J. Phys.}\ }\textbf {\bibinfo {volume} {30}},\
  \bibinfo {pages} {23} (\bibinfo {year} {2009})}\BibitemShut {NoStop}%
\bibitem [{\citenamefont {Carrasco}\ and\ \citenamefont
  {Reula}(2016)}]{carrasco2016}%
  \BibitemOpen
  \bibfield  {author} {\bibinfo {author} {\bibfnamefont {F.~L.}\ \bibnamefont
  {Carrasco}}\ and\ \bibinfo {author} {\bibfnamefont {O.~A.}\ \bibnamefont
  {Reula}},\ }\href {\doibase 10.1103/PhysRevD.93.085013} {\bibfield  {journal}
  {\bibinfo  {journal} {Phys. Rev. D}\ }\textbf {\bibinfo {volume} {93}},\
  \bibinfo {pages} {085013} (\bibinfo {year} {2016})}\BibitemShut {NoStop}%
\bibitem [{\citenamefont {Gill}\ and\ \citenamefont
  {Zachary}(2011)}]{tepper2011}%
  \BibitemOpen
  \bibfield  {author} {\bibinfo {author} {\bibfnamefont {T.}~\bibnamefont
  {Gill}}\ and\ \bibinfo {author} {\bibfnamefont {W.}~\bibnamefont {Zachary}},\
  }\href {\doibase 10.1007/s10701-009-9331-8} {\bibfield  {journal} {\bibinfo
  {journal} {Found. Phys.}\ }\textbf {\bibinfo {volume} {41}},\ \bibinfo
  {pages} {99} (\bibinfo {year} {2011})}\BibitemShut {NoStop}%
\bibitem [{\citenamefont {Noether}(1918)}]{noether1918}%
  \BibitemOpen
  \bibfield  {author} {\bibinfo {author} {\bibfnamefont {E.}~\bibnamefont
  {Noether}},\ }\href@noop {} {\bibfield  {journal} {\bibinfo  {journal}
  {König. Gesellsch. d. Wiss. zu Göttingen, Math.-Phys. Klasse}\ ,\ \bibinfo
  {pages} {235}} (\bibinfo {year} {1918})}\BibitemShut {NoStop}%
\bibitem [{\citenamefont {Kholmetskii}\ \emph {et~al.}(2016)\citenamefont
  {Kholmetskii}, \citenamefont {Missevitch},\ and\ \citenamefont
  {Yarman}}]{kholmetskii2016}%
  \BibitemOpen
  \bibfield  {author} {\bibinfo {author} {\bibfnamefont {A.}~\bibnamefont
  {Kholmetskii}}, \bibinfo {author} {\bibfnamefont {O.}~\bibnamefont
  {Missevitch}}, \ and\ \bibinfo {author} {\bibfnamefont {T.}~\bibnamefont
  {Yarman}},\ }\href {\doibase 10.1007/s10701-015-9963-9} {\bibfield  {journal}
  {\bibinfo  {journal} {Found. Phys.}\ }\textbf {\bibinfo {volume} {46}},\
  \bibinfo {pages} {236} (\bibinfo {year} {2016})}\BibitemShut {NoStop}%
\bibitem [{\citenamefont {Charap}(2011)}]{charap2011}%
  \BibitemOpen
  \bibfield  {author} {\bibinfo {author} {\bibfnamefont {J.}~\bibnamefont
  {Charap}},\ }\href {https://books.google.ca/books?id=H1WPjk\_QIGkC} {\emph
  {\bibinfo {title} {Covariant Electrodynamics: A Concise Guide}}}\ (\bibinfo
  {publisher} {JHUP},\ \bibinfo {year} {2011})\BibitemShut {NoStop}%
\bibitem [{\citenamefont {Schust}\ \emph {et~al.}(2005)\citenamefont {Schust},
  \citenamefont {Stary}, \citenamefont {Mattes},\ and\ \citenamefont
  {Sorg}}]{schust2005}%
  \BibitemOpen
  \bibfield  {author} {\bibinfo {author} {\bibfnamefont {P.}~\bibnamefont
  {Schust}}, \bibinfo {author} {\bibfnamefont {F.}~\bibnamefont {Stary}},
  \bibinfo {author} {\bibfnamefont {M.}~\bibnamefont {Mattes}}, \ and\ \bibinfo
  {author} {\bibfnamefont {M.}~\bibnamefont {Sorg}},\ }\href {\doibase
  10.1007/s10701-005-5830-4} {\bibfield  {journal} {\bibinfo  {journal} {Found.
  Phys.}\ }\textbf {\bibinfo {volume} {35}},\ \bibinfo {pages} {1043} (\bibinfo
  {year} {2005})}\BibitemShut {NoStop}%
\bibitem [{\citenamefont {Bak}\ \emph {et~al.}(1994)\citenamefont {Bak},
  \citenamefont {Cangemi},\ and\ \citenamefont {Jackiw}}]{jackiw1994}%
  \BibitemOpen
  \bibfield  {author} {\bibinfo {author} {\bibfnamefont {D.}~\bibnamefont
  {Bak}}, \bibinfo {author} {\bibfnamefont {D.}~\bibnamefont {Cangemi}}, \ and\
  \bibinfo {author} {\bibfnamefont {R.}~\bibnamefont {Jackiw}},\ }\href
  {\doibase 10.1103/PhysRevD.49.5173} {\bibfield  {journal} {\bibinfo
  {journal} {Phys. Rev. D}\ }\textbf {\bibinfo {volume} {49}},\ \bibinfo
  {pages} {5173} (\bibinfo {year} {1994})}\BibitemShut {NoStop}%
\bibitem [{\citenamefont {Sundermeyer}(1982)}]{sundermeyer1982}%
  \BibitemOpen
  \bibfield  {author} {\bibinfo {author} {\bibfnamefont {K.}~\bibnamefont
  {Sundermeyer}},\ }\href {https://books.google.ca/books?id=R3AzmQEACAAJ}
  {\emph {\bibinfo {title} {Constrained Dynamics}}}\ (\bibinfo  {publisher}
  {Springer-Verlag},\ \bibinfo {year} {1982})\BibitemShut {NoStop}%
\bibitem [{\citenamefont {Belinfante}(1940)}]{belinfante1940}%
  \BibitemOpen
  \bibfield  {author} {\bibinfo {author} {\bibfnamefont {F.}~\bibnamefont
  {Belinfante}},\ }\href {\doibase
  http://dx.doi.org/10.1016/S0031-8914(40)90091-X} {\bibfield  {journal}
  {\bibinfo  {journal} {Physica}\ }\textbf {\bibinfo {volume} {7}},\ \bibinfo
  {pages} {449 } (\bibinfo {year} {1940})}\BibitemShut {NoStop}%
\bibitem [{\citenamefont {Misner}\ \emph {et~al.}(1973)\citenamefont {Misner},
  \citenamefont {Thorne},\ and\ \citenamefont
  {Wheeler}}]{misner1973gravitation}%
  \BibitemOpen
  \bibfield  {author} {\bibinfo {author} {\bibfnamefont {C.}~\bibnamefont
  {Misner}}, \bibinfo {author} {\bibfnamefont {K.}~\bibnamefont {Thorne}}, \
  and\ \bibinfo {author} {\bibfnamefont {J.}~\bibnamefont {Wheeler}},\ }\href
  {https://books.google.ca/books?id=w4Gigq3tY1kC} {\emph {\bibinfo {title}
  {Gravitation}}}\ (\bibinfo  {publisher} {W. H. Freeman},\ \bibinfo {year}
  {1973})\BibitemShut {NoStop}%
\bibitem [{\citenamefont {Eriksen}\ and\ \citenamefont
  {Leinaas}(1980)}]{eriksen1980}%
  \BibitemOpen
  \bibfield  {author} {\bibinfo {author} {\bibfnamefont {E.}~\bibnamefont
  {Eriksen}}\ and\ \bibinfo {author} {\bibfnamefont {J.~M.}\ \bibnamefont
  {Leinaas}},\ }\href {http://stacks.iop.org/1402-4896/22/i=3/a=003} {\bibfield
   {journal} {\bibinfo  {journal} {Phys. Scripta}\ }\textbf {\bibinfo {volume}
  {22}},\ \bibinfo {pages} {199} (\bibinfo {year} {1980})}\BibitemShut
  {NoStop}%
\bibitem [{\citenamefont {Burgess}(2002)}]{burgess2002}%
  \BibitemOpen
  \bibfield  {author} {\bibinfo {author} {\bibfnamefont {M.}~\bibnamefont
  {Burgess}},\ }\href {https://books.google.ca/books?id=er47v803fq0C} {\emph
  {\bibinfo {title} {Classical Covariant Fields}}},\ Cambridge Monogr. Math.
  Phys.\ (\bibinfo  {publisher} {CUP},\ \bibinfo {year} {2002})\BibitemShut
  {NoStop}%
\end{thebibliography}%

\end{document}